\documentclass[12pt]{article}
\usepackage[a4paper]{geometry}
\usepackage{amsfonts,amsmath,amssymb}
\usepackage{mathrsfs}
\usepackage{color}
\usepackage{theorem,cite}

\newtheorem{thm}{Theorem}[section]

\newtheorem{lemma}[thm]{Lemma}
\newtheorem{cor}[thm]{Corollary}

\theorembodyfont{\rmfamily}
\newtheorem{rmk}{Remark}[section]

\newcommand{\bea}{\begin{eqnarray}}
\newcommand{\eea}{\end{eqnarray}}
\newcommand{\beali}{\begin{align}}
\newcommand{\eeali}{\end{align}}
\newcommand{\beano}{\begin{eqnarray*}}
\newcommand{\eeano}{\end{eqnarray*}}
\newcommand{\beq}{\begin{equation}}
\newcommand{\eeq}{\end{equation}}

\newcommand{\CC}{{\mathbb C}}

\newcommand{\II}{{\mathbb I}}

\newcommand{\ZZ}{{\mathbb Z}}

\def\fg{{\mathfrak g}}

\newcommand{\wh}[1]{\widehat{#1}}

\newcommand{\half}{{\textstyle{\frac{1}{2}}}}

\newcommand{\elpa}[1]{{\mathcal{A}_{q,p}(\widehat{gl}_{#1})_c}}
\newcommand{\elpbg}{{\mathcal{B}_{q,\lambda}(\hat{\mathfrak{g}})}}
\newcommand{\elpb}[1]{{{\mathcal{B}}_{q,\lambda}(\widehat{gl}_{#1})_c}}

\newcommand{\proof}{\textbf{Proof:}~}
\newcommand{\qed}{{\hfill \rule{5pt}{5pt}}\\}
\newcommand{\bb}{{\bar b}}
\newcommand{\bc}{{\bar c}}

\def\rank{\mbox{\rm rank}\,}

\def\End{\mathop{\rm End}\nolimits}

\def\id{\mathop{\rm id}\nolimits}
\def\tr{\mathop{\rm tr}\nolimits}

\numberwithin{equation}{section}

\begin{document}
\pagestyle{empty}

\null
\vspace{20pt}

\parskip=6pt

\begin{center}

\begin{LARGE}
\textbf{Dynamical centers for the elliptic \\ quantum algebra $\elpb{2}$}
\end{LARGE}

\vspace{50pt}

\begin{large}
{J.~Avan${}^{a}$, L.~Frappat${}^b$, E.~Ragoucy${}^b$ \footnote[1]{avan@u-cergy.fr, luc.frappat@lapth.cnrs.fr, eric.ragoucy@lapth.cnrs.fr}}
\end{large}

\vspace{15mm}

${}^a$ \textit{Laboratoire de Physique Th\'eorique et Mod\'elisation (CNRS UMR 8089),} \\
\textit{Universit\'e de Cergy-Pontoise, F-95302 Cergy-Pontoise, France} \\

\vspace{5mm}

${}^b$ \textit{Laboratoire d'Annecy-le-Vieux de Physique Th\'eorique LAPTh,} \\
\textit{CNRS -- Universit\'e Savoie Mont Blanc,} \\
\textit{BP 110 Annecy-le-Vieux, F-74941 Annecy cedex, France}

\end{center}

\vspace{4mm}

\begin{abstract}
We identify generating functionals that satisfy dynamical exchange relations with the Lax matrices defining the face-type elliptic quantum algebra $\elpb{2}$, when the central charge takes the two possible values $c =\pm 2$. These generating functionals are constructed as quadratic trace-like objects in terms of the Lax matrices. The obtained structures are characterized as ``dynamical centers'', i.e. the centrality property is deformed by dynamical shifts. For these values, the functionals define (genuine) abelian subalgebras of $\elpb{2}$.

\end{abstract}

\vspace{10mm}
\begin{center}
\emph{Dedicated to the memory of Petr Petrovich Kulish}
\end{center}
\vspace{10mm}

\vfill
\rightline{LAPTH-003/17\qquad\qquad}
\vfill

\newpage

\baselineskip=16.5pt
\parskip=6pt
\parindent=0pt
\pagestyle{plain}
\setcounter{page}{1}

\section{Introduction}

Deformed Virasoro algebras (DVA) have been the subject of many investigations due to their arising in various contexts: relation with Macdonald polynomials \cite{AKOS,SKAO}; symmetries of RSOS models \cite{LP}; AGT duality extended to 5D superconformal gauge theories \cite{AGT,AwaYam}; ZF algebras in XYZ spin chains \cite{Luk96}. It is therefore an interesting question to define algebraic constructions of quadratic exchange structures reproducing the original DVA algebra \cite{SKAO} or generalizing it to situations with more parameters or different deformations of exchange relations. A natural setting to get such constructions was to examine the quadratic trace-like objects built from Lax matrices for elliptic quantum algebras $\elpa{N}$, by analogy with the well-known construction in \cite{FR1995} of extended centers of quantum affine algebras carrying a natural deformed Virasoro Poisson structure.

In a previous paper \cite{AFR} we have proposed a construction of Deformed Virasoro Algebra, more specifically within the context of two-parameter elliptic deformation originally exemplified in \cite{SKAO}. 
Our construction used the quantum Lax matrix $L(z)$ encapsulating the generators of the elliptic affine quantum algebra $\elpa{N}$ \cite{FIJKMY,FIJKMY95,AFRS99}, defined by the famous quadratic exchange Yang--Baxter relation
\begin{equation}
R_{12}(z_1/z_2) \, L_1(z_1) \, L_2(z_2) = L_2(z_2) \, L_1(z_1) \, R_{12}^*(z_1/z_2)
\label{eq:RLLAqp}
\end{equation} 
Here $R_{12}(z)$ is the vertex elliptic quantum $R$-matrix \cite{Bela,ChCh}, $R^{*}_{12}(z) = R_{12}(z)\vert_{p \to p^*=pq^{-2c}}$ and $p$ is the elliptic nome. A set of integer-labeled DVA's was obtained. 
Their generating functionals took the form of a quadratic trace in terms of $L$. 
It paralleled the construction of quantum Virasoro algebra from quantum current algebras pioneered in e.g. \cite{RSTS}.
It was possible to identify both quantum and classical (Poisson) structures for such objects, characterized by the successive implementation of two algebraic conditions, parametrized by two integers, on the algebra parameters $N$, $p$ (elliptic nome), $q$ (deformation parameter) and $c$ (central charge). 
In addition, two distinguished values $c=\pm N$ were identified, where the quadratic object becomes a generating functional of an extended center.

We propose to extend this analysis to the case of dynamical elliptic quantum algebras $\elpbg$, restricting ourselves in a first step to the algebra $\fg=gl_2$.
Dynamical quantum algebras have been subject to many discussions and the literature on this is very large.
It includes (specifically to our purpose) several key results on the quasi-Hopf structure \cite{Drin1990} and Drinfel'd twist interpretation \cite{JKOS,ABRR,BRT}.
It must be emphasized here that the denomination of ``dynamical quantum algebras'' is ambiguous and covers a number of distinct algebraic structures.
Their common key feature however is their interpretation as deformations of Hopf algebra structures, where the deformation parameters are identified as coordinates, generically denoted $\lambda$, on the dual of a classical (Lie or affine) algebra structure.
In the case considered here, this structure is in fact the Cartan subalgebra of an affine algebra.
(Non abelian linear structures yield e.g. so-called non abelian dynamical algebras, see \cite{Ping}).
It follows in particular that such deformed (quasi-Hopf) algebras naturally feature deformation both by coordinates on the dual $\bar{\mathfrak{h}}^*$ 
of the Cartan subalgebra $\bar{\mathfrak{h}}$ of the finite part of the affine algebra, and by a coordinate along the dual central charge $c^*$, identified with $\frac12\ln p/\ln q$ and yielding in a natural way \emph{elliptic} structure functions \cite{JKOS}. Coordinates on $\bar{\mathfrak{h}}^*$ are usually characterized as dynamical parameters.

Note that promoting these dynamical parameters as actual quantum operators by adding explicitly their conjugated momenta 
 to $\elpb{N}$, one gets another dynamical algebra $U_{q,p}(\widehat{gl}_N)_c$ with a very distinct structure, studied in e.g. \cite{Konno,JKOS2,KoKo,Konno2}.

Our purpose is to study whether quadratic trace-like objects, in terms of the Lax matrix generators of $\elpb{2}$, may exhibit similar closure/centrality/abelianity relations as were identified in \cite{AFR} for $\elpa{N}$. 
As in the non dynamical elliptic situation we identify two values $c=\pm 2$ where a quadratic trace-like generating function may be built. 
It obeys the following exchange algebra with the generators of $\elpb{2}$ encapsulated in two quantum Lax matrices $L^{\pm}(z',\lambda)$ ($z'$ is the spectral parameter):
\begin{alignat}{2}
t(z,\lambda) \; L^{\pm}(z',\lambda) &= L^{\pm}(z',\lambda) \; t(z,\lambda+\sigma_z) & \qquad &\text{for $c=-2$} 
\label{eq:tL} \\
t^*(z,\lambda) \; L^{\pm}(z',\lambda) &= L^{\pm}(z',\lambda) \; t^*(z,\lambda+\sigma_z) & \qquad &\text{for $c=2$}
\label{eq:tstL}
\end{alignat}
The matrices $L^{\pm}(z',\lambda)$ live in $\End V \otimes \elpb{2}$ where $V$ is a $2$-dimensional vector space, in other words $L^{\pm}(z',\lambda)$ are $2 \times 2$ matrices of algebra generators. 
$V$ is a weight space of the finite Cartan subalgebra $\bar{\mathfrak{h}}$ which acts on it by $\sigma_z$ (here $\bar{\mathfrak{h}}$ is one-dimensional).

Equations \eqref{eq:tL}--\eqref{eq:tstL} characterize the generating functionals $t(z,\lambda)$ (or $t^*(z,\lambda)$) as what we call ``dynamical center''.
Precise understanding of this property is yet lacking.
In addition, $t(z,\lambda)$ (resp. $t^*(z,\lambda)$) generates, when $c=-2$ (resp. when $c=2$), an abelian subalgebra of $\elpb{2}$.
The notation $t^*(z,\lambda)$ reflects here and throughout the paper a second quadratic trace-like expression for a consistent generating functional. It is \emph{not} related to the notation $p^*=pq^{-2c}$, hopefully there will be no confusion.
A subtle issue arises of which local completion of the enveloping algebra $\elpb{2}$ the generating functionals belong to. It is discussed in detail in section \ref{sect:comment}. 

The paper runs as follows. In Section 2 we prepare some definitions and meaningful properties of the elliptic dynamical algebra $\elpbg$. 
In particular, when $\wh{\mathfrak{g}}=\wh{gl}_2$, we identify a ``crossing'' relation connecting partially transposed quantum $R$-matrices, which plays a crucial role here, there and everywhere. 
The main results are stated and proved in Section 3.
Finally, Section 4 contains some conclusive remarks on open questions and possible follow-ups.

\section{The dynamical elliptic quantum algebra $\elpbg$}

\subsection{Definition of $\elpbg$}

The face-type or dynamical elliptic quantum algebra $\elpbg$ is a quasi-triangular quasi-Hopf algebra obtained by a Drinfeld twist procedure on the quantum group $\mathcal{U}_q(\hat{\mathfrak{g}})$, where $\hat{\mathfrak{g}}$ is an affine Kac--Moody algebra \cite{Fron,JKOS}.
More precisely, the quantum group $\mathcal{U}_q(\hat{\mathfrak{g}})$ is a quasi-triangular Hopf algebra equipped with a coproduct $\Delta$, counit $\varepsilon$, antipode $S$ and universal $R$-matrix $\mathcal{R}$.
If $\mathfrak{h} = \bar{\mathfrak{h}} \oplus c \oplus d$ is a Cartan subalgebra of $\hat{\mathfrak{g}}$, we denote by  
$\{h_l\}$ a standard basis of $\bar{\mathfrak{h}}$ and $\{h^l\}$ its dual basis $(l=1,2,...,\rank \mathfrak{g})$.
We consider the face-type twistor $\mathcal{F}(\lambda)$ \cite{JKOS}, the invertible element in $\mathcal{U}_q(\hat{\mathfrak{g}}) \otimes \mathcal{U}_q(\hat{\mathfrak{g}})$ depending on $\lambda \in \mathfrak{h}$, such that
\begin{equation}
(\id \otimes \, \varepsilon) \mathcal{F}(\lambda) = (\varepsilon \otimes \id) \mathcal{F}(\lambda) = 1
\end{equation}
and
\begin{equation}
\mathcal{F}^{(12)}(\lambda) (\Delta \otimes \id) \mathcal{F}(\lambda) = \mathcal{F}^{(23)}(\lambda+h^{(1)}) (\id \otimes \, \Delta) \mathcal{F}(\lambda) \,.
\end{equation}
This last equation is the so-called shifted cocycle condition.

The dynamical elliptic quantum algebra $\elpbg$ is a quasi-triangular quasi-Hopf algebra with twisted universal $R$-matrix
\begin{equation}
\mathcal{R}(\lambda) = \mathcal{F}^{(21)}(\lambda) \, \mathcal{R} \, \mathcal{F}^{(12)}(\lambda)^{-1}
\end{equation}
twisted coproduct
\begin{equation}
\Delta_\lambda(x) = \mathcal{F}(\lambda) \Delta(x) \mathcal{F}(\lambda) \qquad \forall x \in \mathcal{U}_q(\hat{\mathfrak{g}})
\end{equation}
and coassociator
\begin{equation}
\Phi(\lambda) = \mathcal{F}^{(23)}(\lambda) \, \mathcal{F}^{(23)}(\lambda+h^{(1)})^{-1} \,.
\end{equation}
The universal $R$-matrix $\mathcal{R}(\lambda)$ satisfies the so-called \emph{Gervais--Neveu--Felder} or 
\emph{dynamical Yang--Baxter equation} \cite{GNF}:
\begin{equation}
\mathcal{R}_{12}(\lambda + h^{(3)}) \, \mathcal{R}_{13}(\lambda) \, \mathcal{R}_{23}(\lambda + h^{(1)}) 
= \mathcal{R}_{23}(\lambda) \, 
\mathcal{R}_{13}(\lambda + h^{(2)}) \, 
\mathcal{R}_{12}(\lambda).
\label{eq:DYBE}
\end{equation}
This equation is written in $\elpbg\otimes \elpbg\otimes \elpbg$.
The notation $\lambda+h^{(j)}$ denotes a shift of  $\lambda$ by the Cartan generator $\sum_{l=1}^{N} h_{l}$
acting in the $j^{th}$ copy of $\elpbg$.
We set
$\lambda = \rho + \sum_{l=1}^{N} \lambda^{l} h_{l} + (r+N)d + \xi c$ 
where $\rho$ is the Weyl vector and $\lambda^l, r, \xi \in \CC$. 

It must be emphasized that this universal R-matrix obeys a \textit{zero-weight condition} 
\begin{equation}
[h^{(1)}+h^{(2)}, \mathcal{R}(\lambda)] = 0 
\label{eq:zeroweight}
\end{equation}
due to the facts that the initial $R$-matrix already obeys it by construction
and the face-type twist element $\mathcal{F}(\lambda)$ also obeys it by construction (see \cite{JKOS}). 
Subsequent zero-weight conditions will follow from application of suitable representation morphisms.
The zero-weight condition formulated  in \eqref{eq:zeroweight} is extendable
to any matrix with the suitable tensor structure $M_{12}$.

In order to obtain the dynamical RLL relations, one introduces the following matrices:
\begin{align}
\begin{split}
& \mathcal{R}^+(\lambda) = q^{c \otimes d + d \otimes c} \, \mathcal{R}(\lambda), \\
& \mathcal{R}^-(\lambda) = \mathcal{R}_{21}(\lambda)^{-1} \, q^{-c \otimes d - d \otimes c}.
\end{split}
\end{align}
Denoting by $\pi_{V}(z)$ an evaluation representation of $\mathcal{U}_q(\hat{\mathfrak{g}})$ 
with evaluation parameter $z$ and representation space $V$, where $V$ is a weight space of $\bar{\mathfrak{h}}$, the Lax matrices 
$L^\pm(z,\lambda) = (\pi_{V}(z) \otimes \id) \, \mathcal{R}^\pm(\lambda)$ 
realize an FRT-type formalism of $\elpbg$ with an evaluated $R$-matrix defined by 
$R(z_1/z_2,\lambda) = \big( \pi_{V}(z_1) \otimes \pi_{V}(z_2) \big) \, \mathcal{R}(\lambda)$. 
The RLL relations take the form
\begin{align}
\begin{split}
& R_{12}(z_1/z_2,\lambda+h) \, L^\pm_1(z_1,\lambda) \, L^\pm_2(z_2,\lambda+h^{(1)}) = 
L^\pm_2(z_2,\lambda) \, L^\pm_1(z_1,\lambda+h^{(2)}) \, R_{12}(z_1/z_2,\lambda)\,, \\
& R_{12}(q^c z_1/z_2,\lambda+h) \, L^+_1(z_1,\lambda) \, L^-_2(z_2,\lambda+h^{(1)}) = 
L^-_2(z_2,\lambda) \, L^+_1(z_1,\lambda+h^{(2)}) \, R_{12}(q^{-c} z_1/z_2,\lambda) .
\end{split}
\label{eq:RLL}
\end{align}
$h^{(1)}$ and $h^{(2)}$ are the evaluation of the Cartan generators $h$ in the first and second copies of $V$ respectively.
Note that the operators $L^\pm$ are not independent, see \cite{JKOS}. Equation \eqref{eq:RLL} contains expressions of shifts for the whole set of deformation parameters $\lambda$ by suitable Cartan generators. In particular, the elliptic nome $p$ introduced as $p=q^{2r}$ in $R_{12}$ is shifted as $pq^{-2c}$ where $c$ is the central charge generator in the abstract algebra $\hat{\mathfrak{g}}$. This yields the $p \to p^*$ shift whenever a value is assigned to $c$ by a choice of representation of $\elpbg$.

\subsection{$R$-matrix of $\elpb{2}$ in the fundamental representation}

In the following, we restrict ourselves to the case $\hat{\mathfrak{g}} = \widehat{gl}_2$ and we consider the two-dimensional evaluation representation ($V=\CC^2$).
The evaluated $R$-matrix of $\elpb{2}$ takes the following form \cite{GNF}:
\begin{equation}
R(z,\lambda) = \rho(z) 
\begin{pmatrix}
1 & 0 & 0 & 0 \\
0 & b(z) & c(z) & 0 \\
0 & \bc(z) & \bb(z) & 0 \\
0 & 0 & 0 & 1 \\
\end{pmatrix} 
\label{eq:Rbqp}
\end{equation}
where
\begin{alignat}{2}
& b(z) = \frac{\Theta_{p}(q^2 w) \, \Theta_{p}(z)}{\Theta_{p}(w) \, \Theta_{p}(q^2 z)} \;,
&\qquad& \bb(z) = \frac{\Theta_{p}(q^2 w^{-1}) \, \Theta_{p}(z)}{\Theta_{p}(w^{-1}) \, \Theta_{p}(q^2 z)} \;, \\
& c(z) = \frac{\Theta_{p}(q^2) \, \Theta_{p}(wz)}{\Theta_{p}(w) \, \Theta_{p}(q^2z)} \;, 
&\qquad& \bc(z) = \frac{\Theta_{p}(q^2) \, \Theta_{p}(w^{-1}z)}{\Theta_{p}(w^{-1}) \, \Theta_{p}(q^2z)} \;.
\label{def:ccbar}
\end{alignat}
Here we denote $\lambda = \rho + \half\,s\, \sigma_z + (r+2)d + \xi c$. 
The dynamical parameter $w$ is related to the deformation parameter $q$ by $w=q^{2s}$ 
and $z$ is the spectral parameter.

We recall that the Jacobi $\Theta$ function is defined by
$\Theta_p(z) = (z;p)_\infty \, (pz^{-1};p)_\infty \, (p;p)_\infty$
where the infinite multiple Pochhammer products are given by
$(z;p_1,\dots,p_m)_\infty = \prod_{n_i \ge 0} (1-z p_1^{n_1} \dots p_m^{n_m})$.
The $\Theta_{p}(z)$ function enjoys the following property: $\Theta_{p}(pz) = \Theta_{p}(z^{-1}) = -z^{-1}\,\Theta_{p}(z)$. \\
The normalization factor $\rho(z)$ is given by
\begin{equation}
\rho(z) = q^{-1/2} \, \frac{(q^2z;p,q^4)_{\infty}^2 \; (pz^{-1};p,q^4)_{\infty} \; (pq^4z^{-1};p,q^4)_{\infty}}{(pq^2z^{-1};p,q^4)_{\infty}^2 \; (z;p,q^4)_{\infty} \; (q^4z;p,q^4)_{\infty}} \;.
\label{eq:rhoelpb}
\end{equation}
The $R$-matrix of $\elpb{2}$ used in \cite{JKOS} is obtained from (\ref{eq:Rbqp}) by a harmless twist transformation:
\begin{equation}
\widetilde R_{12}(z,\lambda) = g_2(\lambda) \, g_1(\lambda+h^{(2)}) \, R_{12}(z,\lambda) \, g_1^{-1}(\lambda) \, g_2^{-1}(\lambda+h^{(1)})
\label{eq:twist}
\end{equation}
where $g$ is a spectral parameter independent diagonal matrix with $g_{11}(\lambda)=1$ and $g_{22}(\lambda) = w^{-1/2} (w;p)_{\infty} \, (pq^2w^{-1};p)_{\infty}$. 
Explicitly, it reads
\begin{equation}
\widetilde R(z,\lambda) = \rho(z) 
\begin{pmatrix}
1 & 0 & 0 & 0 \\
0 & b'(z) & c(z) & 0 \\
0 & \bc(z) & \bb'(z) & 0 \\
0 & 0 & 0 & 1 \\
\end{pmatrix} 
\label{eq:Rbqpalt}
\end{equation}
where $c(z)$, $\bc(z)$, $\rho(z)$ are given by \eqref{def:ccbar}, \eqref{eq:rhoelpb} and
\begin{align}
b'(z) &= q \, \frac{(pw^{-1}q^2;p)_{\infty} \; (pw^{-1}q^{-2};p)_{\infty}}
{(pw^{-1};p)_{\infty}^2} \; \frac{\Theta_{p}(z)}{\Theta_{p}(q^2z)} \;, \\
\bb'(z) &= q \, \frac{(wq^2;p)_{\infty} \; (wq^{-2};p)_{\infty}}
{(w;p)_{\infty}^2} \; \frac{\Theta_{p}(z)}{\Theta_{p}(q^2z)} \;.
\end{align}

Now, the RLL relations in \cite{JKOS} read
\begin{align}
\begin{split}
& \widetilde R_{12}(z_1/z_2,\lambda+h) \, \widetilde L^\pm_1(z_1,\lambda) \, \widetilde L^\pm_2(z_2,\lambda+\sigma_z^{(1)}) = 
\widetilde L^\pm_2(z_2,\lambda) \, \widetilde L^\pm_1(z_1,\lambda+\sigma_z^{(2)}) \, \widetilde R_{12}(z_1/z_2,\lambda) \\
& \widetilde R_{12}(q^c z_1/z_2,\lambda+h) \, \widetilde L^+_1(z_1,\lambda) \, \widetilde L^-_2(z_2,\lambda+\sigma_z^{(1)}) = 
\widetilde L^-_2(z_2,\lambda) \, \widetilde L^+_1(z_1,\lambda+\sigma_z^{(2)}) \, \widetilde R_{12}(q^{-c} z_1/z_2,\lambda) 
\end{split}
\label{eq:RLLalt}
\end{align}
where
\begin{equation}
\widetilde L^\pm(z,\lambda) = g(\lambda+\sigma_z) \, L^\pm(z,\lambda) \, g(\lambda)^{-1}
\label{eq:laxalt}
\end{equation}
Here we assume, as usually done, that $\big[ \sigma_z+h , L^\pm(z,\lambda) \big] = 0$.

The matrices \eqref{eq:Rbqp} and \eqref{eq:Rbqpalt} satisfy the following unitary condition:
\begin{equation}
R_{12}(z,\lambda) R_{21}(z^{-1},\lambda) = \widetilde R_{12}(z,\lambda) \widetilde R_{21}(z^{-1},\lambda) = \rho(z) \, \rho(z^{-1})
\label{eq:unitair}
\end{equation}
For simplicity, we set $\mathfrak{n}(z) = \rho(z) \, \rho(z^{-1})$. Note that the function $\mathfrak{n}(z)$ is $q^4$-periodic.

A crucial property relates the $R$-matrix to its partial transposition. It plays the role of the crossing relation for non dynamical elliptic matrices. It is given by the following theorem:
\begin{thm}
\label{thmcross}
The dynamical elliptic $R$-matrices \eqref{eq:Rbqp} and \eqref{eq:Rbqpalt} satisfy the following crossing relations:
\begin{equation}
\sigma_y^{(1)} \left( R_{12}^{t_1}(z^{-1}q^{-2},\lambda)\right)^{-sl_1}
\sigma_y^{(1)} \; \frac{\Upsilon(\lambda+\sigma_z^{(2)})}{\Upsilon(\lambda)}
= R_{12}^{-1}(z^{-1},\lambda)
\label{eq:cross}
\end{equation}
and
\begin{equation}
\sigma_y^{(1)} \, \Gamma_1(\lambda) \big( \widetilde R_{12}^{t_1}(z^{-1}q^{-2},\lambda) \big)^{-sl_1}
\Gamma_1(\lambda+\sigma_z^{(2)})^{-1} \, \sigma_y^{(1)} \; \frac{\Upsilon(\lambda+\sigma_z^{(2)})}{\Upsilon(\lambda)}
= \widetilde R_{12}^{-1}(z^{-1},\lambda)
\label{eq:crossalt}
\end{equation}
where $\Upsilon(\lambda) = w^{-1/2} \, \Theta_p(w)$ and $\Gamma(\lambda) = (\det g) \, g(\lambda)^{-1} \, g(\lambda)^{-sc}$.

For any matrix $M$, we remind the shift-column ($sc$) and shift-line [or shift-row] ($sl$) notations, defined as follows:
\begin{align}
& \big( M \big)^{sc} = \big( e^{\sigma_z \partial} M^{t} \big)^{t} \, e^{-\sigma_z \partial}
 = \big( e^{\sigma_z \partial} \big( M\, e^{-\sigma_z \partial} \big)^{t} \big)^{t} \nonumber \\
& \big( M \big)^{sl} = \big( \big( e^{\sigma_z \partial} M\big)^{t} \, e^{-\sigma_z \partial} \big)^{t}
= e^{\sigma_z \partial} \big( M^{t} \, e^{-\sigma_z \partial} \big)^{t},
\end{align}
where $\partial$ denotes $\dfrac{\partial}{\partial\lambda}$.
If $M_{12}$ lies in a tensor space $\text{End}(V)\otimes\text{End}(V)$, the shift-column $sc_1$, $sc_2$
and shift-line $sl_1$, $sl_2$ notations are defined accordingly.
\end{thm}
\proof Relation \eqref{eq:cross} has been proved in \cite{Filali} and relation \eqref{eq:crossalt} is a consequence of \eqref{eq:cross} and the twist transformation \eqref{eq:twist}. 
We used the relation 
$\big( M_{12}^{t_1} \big)^{sc_1} = \big( M_{12}^{sl_1} \big)^{t_1}$. Remark that although $\Upsilon(\lambda)$ is a scalar function, the shift in $\Upsilon(\lambda+\sigma_z^{(2)})$ promotes it to a genuine matrix.
\qed

\begin{cor}\label{cor22}
The crossing-unitarity relation \cite{JMO}
\begin{equation}
\left( \big( \widetilde R_{12}(z^{-1}q^{-4},\lambda)^{-sl_2}\big)^{t_1} \right)^{-1} = \frac{1}{\mathfrak{n}(z)} \, G_1(\lambda)^{-1} \, \big( \widetilde R_{21}^{t_1}(z,\lambda)\big)^{-sc_2} \, G_1(\lambda-\sigma_z^{(2)}) \,,
\label{eq:cor22}
\end{equation}
where $G(\lambda) = \dfrac{\Upsilon(\lambda)}{\Upsilon(\lambda+\sigma_z)}$\,,
obeyed by the $R$-matrix \eqref{eq:Rbqpalt} of the dynamical elliptic quantum algebra $\elpb{2}$, can be deduced from the crossing relation of Theorem \ref{thmcross}. \\
The $R$-matrix \eqref{eq:Rbqp} also satisfies \eqref{eq:cor22}.
\end{cor}

\proof 
The crossing formula \eqref{eq:crossalt} applied twice and the elimination of the $\sigma_y$ matrices will yield \eqref{eq:cor22}.
We start by considering the inverse of \eqref{eq:crossalt}. One obtains
\begin{equation}
\frac{\Upsilon(\lambda)}{\Upsilon(\lambda+\sigma_z^{(2)})} \, \sigma_y^{(1)} \, \Gamma_1(\lambda+\sigma_z^{(2)}) \big( (\widetilde R_{12}^{t_1}(z^{-1}q^{-2},\lambda))^{-sl_1}\big)^{-1} \, \Gamma_1(\lambda)^{-1} \, \sigma_y^{(1)} = \widetilde R_{12}(z^{-1},\lambda) \,.
\label{eq:tmp221}
\end{equation}
Hence one has $\big( (\text{LHS of \eqref{eq:tmp221}})^{-sc_1} \big)^{t_1} = \big( \widetilde R_{12}(z^{-1},\lambda)^{-sc_1}\big)^{t_1} = \big( \widetilde R_{12}^{t_1}(z^{-1},\lambda)\big)^{-sl_1}$ and one uses \eqref{eq:crossalt} to evaluate $\big( \widetilde R_{12}^{t_1}(z^{-1},\lambda)\big)^{-sl_1}$. Therefore one gets, with $z \to zq^2$,
\begin{align}
\frac{\Upsilon(\lambda)}{\Upsilon(\lambda+\sigma_z^{(2)})} \, \sigma_y^{(1)} \, \Big( \Gamma_1(\lambda) \big( \widetilde R_{12}^{t_1}(z^{-1}q^{-4},\lambda) \big)^{-sl_1} \, \Gamma_1(\lambda+\sigma_z^{(2)})^{-1} \Big)^{-1} \, \sigma_y^{(1)} = & \nonumber \\
\left( \left( \Upsilon(\lambda) \, \Gamma_1(\lambda)^{-1} \, \sigma_y^{(1)} \, \widetilde R_{12}^{-1}(z^{-1},\lambda) \, \sigma_y^{(1)} \, \Gamma_1(\lambda+\sigma_z^{(2)}) \, \Upsilon(\lambda+\sigma_z^{(2)})^{-1} \right)^{t_1} \right)^{sc_1} & \,.
\label{eq:tmp222}
\end{align}
We now work out the right hand side of the equation to eliminate the $\sigma_y^{(1)}$ matrices. Due to the property $\Gamma(\lambda)^{-1} \, \sigma_y = (\det g)^{-1}\,(\det g^{-sc})^{-1} \sigma_y \, \Gamma(\lambda)$ and $(\sigma_y^{(1)} \,A_{12}\sigma_y^{(1)})^{t_1} = \sigma_y^{(1)} \,A_{12}^{t_1}\sigma_y^{(1)}$, one has, setting $\mu(\lambda) = (\det g)^{-1} \, (\det g^{-sc})^{-1} \, \Upsilon(\lambda)$,
\begin{equation}
\text{RHS of \eqref{eq:tmp222}} = \left( \mu(\lambda) \, \sigma_y^{(1)} \left( \Gamma_1(\lambda) \, \widetilde R_{12}^{-1}(z^{-1},\lambda) \, \Gamma_1(\lambda+\sigma_z^{(2)})^{-1} \right)^{t_1} \sigma_y^{(1)} \, \mu(\lambda+\sigma_z^{(2)})^{-1} \right)^{sc_1} \,.
\end{equation}
Note that the leftmost $\sigma_y^{(1)}$ matrix does not see the $sc_1$ operation, hence it can be factored out on the left and it simplifies with the LHS. Thus, after moving $\mu(\lambda)$ to the right, Eq. \eqref{eq:tmp222} reads as
\begin{align}
& \Big( \Gamma_1(\lambda) \big( \widetilde R_{12}^{t_1}(z^{-1}q^{-4},\lambda) \big)^{-sl_1} \, \Gamma_1(\lambda+\sigma_z^{(2)})^{-1} \Big)^{-1} \, \sigma_y^{(1)} = \nonumber \\
& \frac{\Upsilon(\lambda+\sigma_z^{(2)})}{\Upsilon(\lambda)} \, \left( \left( \Gamma_1(\lambda) \, \widetilde R_{12}^{-1}(z^{-1},\lambda) \, \Gamma_1(\lambda+\sigma_z^{(2)})^{-1} \right)^{t_1} \sigma_y^{(1)} \right)^{sc_1} \, \frac{\mu(\lambda+\sigma_z^{(1)})}{\mu(\lambda+\sigma_z^{(1)}+\sigma_z^{(2)})} \,.
\label{eq:tmp224}
\end{align}
We now multiply both sides of the equation on the right by $e^{(\sigma_z^{(1)}+\sigma_z^{(2)})\partial}$ and move the exponential to the left. For the left hand side we get:
\begin{align}\label{eq229}
\text{LHS of \eqref{eq:tmp224}} & = \Big( \Gamma_1(\lambda) \big( \widetilde R_{12}^{t_1}(z^{-1}q^{-4},\lambda) \big)^{-sl_1} \, \Gamma_1(\lambda+\sigma_z^{(2)})^{-1} \Big)^{-1} \, e^{(-\sigma_z^{(1)}+\sigma_z^{(2)})\partial} \, \sigma_y^{(1)} \nonumber \\
& = e^{(-\sigma_z^{(1)}+\sigma_z^{(2)})\partial} \left( \big( \Gamma_1(\lambda) \, (\widetilde R_{12}^{t_1}(z^{-1}q^{-4},\lambda))^{-sl_1} \, \Gamma_1(\lambda+\sigma_z^{(2)})^{-1} \big)^{-1} \right)^{sl_1-sl_2} \sigma_y^{(1)} \nonumber \\
& = e^{(-\sigma_z^{(1)}+\sigma_z^{(2)})\partial} \left( \big( \Gamma_1(\lambda) \, (\widetilde R_{12}^{t_1}(z^{-1}q^{-4},\lambda))^{-sl_1} \, \Gamma_1(\lambda+\sigma_z^{(2)})^{-1} \big)^{sl_1-sl_2} \right)^{-1} \sigma_y^{(1)} \,.
\end{align}
The first equality in \eqref{eq229} is due to the relation 
$M_{12} \, e^{(-\sigma_z^{(1)}+\sigma_z^{(2)})\partial} = e^{(-\sigma_z^{(1)}+\sigma_z^{(2)})\partial} \, M_{12}^{sl_1-sl_2}$,
valid for a matrix $M_{12}$ such that $M_{12}^{t_1}$ satisfies the zero-weight condition. 
This last relation implies  that the total shift-line operation $sl_1-sl_2$ and the inversion commute when acting on $M_{12}$, leading to the last equality  in \eqref{eq229}.

A direct computation in components shows that 
\begin{equation}
\big( \Gamma_1(\lambda) \, (\widetilde R_{12}^{t_1}(z^{-1}q^{-4},\lambda))^{-sl_1} \, \Gamma_1(\lambda+\sigma_z^{(2)})^{-1} \big)^{sl_1-sl_2} = \Gamma_1(\lambda-\sigma_z^{(2)})^{sc_1} \big( \widetilde R_{12}^{t_1}(z^{-1}q^{-4},\lambda) \big)^{-sl_2} \, \Gamma_1^{-1}(\lambda)^{sc_1} \,.
\end{equation}
Taking now the inverse, one gets for the LHS of \eqref{eq:tmp224}
\begin{equation}
e^{(-\sigma_z^{(1)}+\sigma_z^{(2)})\partial} \, \Gamma_1(\lambda)^{sc_1} \, \big( \big( \widetilde R_{12}(z^{-1}q^{-4},\lambda)^{-sl_2} \big)^{t_1} \big)^{-1} \, \Gamma_1^{-1}(\lambda-\sigma_z^{(2)})^{sc_1} \, \sigma_y^{(1)} \,.
\label{eq:lhs224}
\end{equation}
To evaluate the right hand side, one applies the following property (proved by a direct calculation in components): if $M_{12}$ is a matrix satisfying the zero-weight condition, then
\begin{equation}
\big( M_{12}^{t_1} \sigma_y^{(1)} \big)^{sc_1} \, e^{(\sigma_z^{(1)}+\sigma_z^{(2)})\partial} = e^{(-\sigma_z^{(1)}+\sigma_z^{(2)})\partial}\big( M_{12}^{t_1} \big)^{-sc_2} \sigma_y^{(1)} \,.
\end{equation}
Taking $M_{12} = \Gamma_1(\lambda) \, \widetilde R_{12}^{-1}(z^{-1},\lambda) \, \Gamma_1(\lambda+\sigma_z^{(2)})^{-1}$, the RHS of \eqref{eq:tmp224} reads as
\begin{equation}
e^{(-\sigma_z^{(1)}+\sigma_z^{(2)})\partial} \frac{\Upsilon(\lambda+\sigma_z^{(1)})}{\Upsilon(\lambda+\sigma_z^{(1)}-\sigma_z^{(2)})} \, \left( \left( \Gamma_1(\lambda) \, \widetilde R_{12}^{-1}(z^{-1},\lambda) \, \Gamma_1(\lambda+\sigma_z^{(2)})^{-1} \right)^{t_1} \right)^{-sc_2} \sigma_y^{(1)} \, \frac{\mu(\lambda-\sigma_z^{(2)})}{\mu(\lambda)}
\label{eq:rhs224}
\end{equation}
Comparing \eqref{eq:lhs224} and \eqref{eq:rhs224}, one obtains
\begin{align}
& \Gamma_1(\lambda)^{sc_1} \, \big( \big( \widetilde R_{12}(z^{-1}q^{-4},\lambda)^{-sl_2} \big)^{t_1} \big)^{-1} \, \Gamma_1^{-1}(\lambda-\sigma_z^{(2)})^{sc_1} \nonumber \\
& = \frac{\Upsilon(\lambda+\sigma_z^{(1)})}{\Upsilon(\lambda+\sigma_z^{(1)}-\sigma_z^{(2)})} \, \left( \left( \Gamma_1(\lambda) \, \widetilde R_{12}^{-1}(z^{-1},\lambda) \, \Gamma_1(\lambda+\sigma_z^{(2)})^{-1} \right)^{t_1} \right)^{-sc_2} \frac{\mu(\lambda-\sigma_z^{(2)})}{\mu(\lambda)}
\label{eq:tmp230}
\end{align}
One has again by a calculation in components and using unitarity:
\begin{equation}
\left( \left( \Gamma_1(\lambda) \, \widetilde R_{12}^{-1}(z^{-1},\lambda) \, \Gamma_1(\lambda+\sigma_z^{(2)})^{-1} \right)^{t_1} \right)^{-sc_2} = \frac{1}{\mathfrak{n}(z)} \, \Gamma_1^{-1}(\lambda) \, \big( \widetilde R_{21}^{t_1}(z,\lambda) \big)^{-sc_2} \, \Gamma_1(\lambda-\sigma_z^{(2)})
\end{equation}
Therefore \eqref{eq:tmp230} is rewritten as
\begin{align}
\big( \widetilde R_{12}(z^{-1}q^{-4},\lambda)^{-sl_2} \big)^{t_1} \big)^{-1} = & \; \frac{1}{\mathfrak{n}(z)} \, \big( \Gamma_1(\lambda)^{sc_1} \big)^{-1} \, \frac{\Upsilon(\lambda+\sigma_z^{(1)})}{\Upsilon(\lambda+\sigma_z^{(1)}-\sigma_z^{(2)})} \, \Gamma_1^{-1}(\lambda) \, \nonumber \\
& \times \big( \widetilde R_{21}^{t_1}(z,\lambda) \big)^{-sc_2} \, \Gamma_1(\lambda-\sigma_z^{(2)}) \, \frac{\mu(\lambda-\sigma_z^{(2)})}{\mu(\lambda)} \, \big( \Gamma_1(\lambda-\sigma_z^{(2)}) \big)^{sc_1}
\end{align}
Since the matrix $\widetilde R_{12}$ satisfies the zero-weight condition, the function $\Upsilon(\lambda+\sigma_z^{(1)}-\sigma_z^{(2)})$ can be sent to the right of $\widetilde R_{21}^{t_1}$ and $\mu(\lambda)$ is sent to the left:
\begin{align}
\big( \widetilde R_{12}(z^{-1}q^{-4},\lambda)^{-sl_2} \big)^{t_1} \big)^{-1} = & \; \frac{1}{\mathfrak{n}(z)} \, \Big( \Gamma_1(\lambda) \, \frac{\mu(\lambda)}{\Upsilon(\lambda+\sigma_z^{(1)})} \, \Gamma_1(\lambda)^{sc_1} \Big)^{-1} \nonumber \\
& \times \big( \widetilde R_{21}^{t_1}(z,\lambda) \big)^{-sc_2} \, \Gamma_1(\lambda-\sigma_z^{(2)}) \, \frac{\mu(\lambda-\sigma_z^{(2)})}{\Upsilon(\lambda+\sigma_z^{(1)}-\sigma_z^{(2)})} \, \big( \Gamma_1(\lambda-\sigma_z^{(2)}) \big)^{sc_1}
\end{align}
We now evaluate $\Gamma_1(\lambda) \, \dfrac{\mu(\lambda)}{\Upsilon(\lambda+\sigma_z^{(1)})} \, \Gamma_1(\lambda)^{sc_1}$ which reduces after an explicit computation to $\dfrac{\Upsilon(\lambda)}{\Upsilon(\lambda+\sigma_z^{(1)})}$. 
We therefore obtain the desired result.

It is seen in the proof that explicit dependence in the twist matrix $g$ drops out by cancellation between the $\mu$ and $G$ contributions. Hence the final crossing property  \eqref{eq:cor22} is twist independent and in particular is also obeyed by the original $R$-matrix \eqref{eq:Rbqp}.
\qed

\section{Dynamical centers in $\elpb{2}$ at distinguished values}

\subsection{Main result}

We now state the main result of the paper, i.e. the existence of ``dynamically central'' quadratic structures in $\elpb{2}$ for the specific values $c=\pm 2$ of the central charge, realizing in addition abelian subalgebras of $\elpb{2}$.
\begin{thm}
\label{thmone}
We introduce the generating functions 
\begin{align}
t(z,\lambda) &= \tr \Big( N(\lambda) \, e^{-\sigma_z\partial} \, {L^+(q^{-c}z,\lambda)}^{-1} \, L^-(z,\lambda) \, e^{\sigma_z\partial} \Big) = \sum_{n \in \ZZ} t_n z^{-n} \,,
\label{eq:deftmn}\\
t^*(z,\lambda) &= \tr \Big( N(\lambda) \, e^{-\sigma_z\partial} \, {L^-(q^{c}z,\lambda)}^{-1} \, L^+(z,\lambda) \, e^{\sigma_z\partial} \Big) = \sum_{n \in \ZZ} t^*_n z^{-n} \,.
\label{eq:deftmnst}
\end{align}
where the diagonal matrix $N(\lambda) \in \End V \otimes \CC(\bar{\mathfrak{h}})$ is given by
\begin{equation}
N(\lambda) = \Big( \frac{\Upsilon(\lambda)}{\Upsilon(\lambda+\sigma_z)} \Big)^{-sc}
= \frac{\Upsilon(\lambda-\sigma_z)}{\Upsilon(\lambda)}
\end{equation}
They obey the exchange relations:
\begin{alignat}{2}
t(z_1,\lambda) \, L^\pm(z_2,\lambda) &= L^\pm(z_2,\lambda) \, t(z_1,\lambda+\sigma_z) & \qquad & \text{when $c=-2$} 
\label{eq:echtL} \\
t^*(z_1,\lambda) \, L^\pm(z_2,\lambda) &= L^\pm(z_2,\lambda) \, t^*(z_1,\lambda+\sigma_z) & \qquad & \text{when $c=2$}
\label{eq:echtstL} 
\end{alignat}
\end{thm}
The proof of this theorem is given in the next subsection.

The fact that the matrix $N(\lambda)$ is diagonal ensures that the generating functionals $t(z_1,\lambda)$ and $t^*(z_1,\lambda)$ lie indeed in the dynamical elliptic quantum algebra $\elpb{2}$ rather than in $U_{q,p}(\widehat{gl}_2)$.

The dynamical shift on $\lambda$ in the exchange relations \eqref{eq:echtL}--\eqref{eq:echtstL} leads us to characterize 
$t(z,\lambda)$ and $t^{*}(z,\lambda)$  as generating functionals for \emph{dynamical} centers in $\elpb{2}$.
Here and below, we will use the notation $t^{(*)}(z,\lambda)$ for relations valid for both $t(z,\lambda)$ and $t^{*}(z,\lambda)$.

The following corollary immediately holds:
\begin{cor}
At distinguished values of the central charge, the generating function $t(z,\lambda)$ (resp. $t^*(z,\lambda)$) realizes an abelian subalgebra of $\elpb{2}$:
\begin{alignat}{2}
&\big[t(z_1,\lambda) \,,\, t(z_2,\lambda)\big] = 0 & \qquad & \text{when $c=-2$} 
 \\
&\big[t^*(z_1,\lambda) \,,\, t^*(z_2,\lambda)\big] = 0 & \qquad & \text{when $c=2$.}
\end{alignat}
\end{cor}
\proof
Rewriting \eqref{eq:echtL}--\eqref{eq:echtstL} as an exact commutation relation $\big[ t^{(*)}(z_1,\lambda) \,,\, L^\pm(z_2,\lambda) \, e^{\sigma_z\partial} \big] = 0$ and noting that $t^{(*)}(z_1,\lambda)$ commutes with the diagonal matrix $N(\lambda)$, one gets the result.
\qed

Inserting the relation \eqref{eq:laxalt} in \eqref{eq:deftmn}--\eqref{eq:deftmnst}, it is immediate to realize that the $g$ dependence drops due to trace cyclicity, hence $t^{(*)}(z,\lambda)$ has the same expression in terms of $L^\pm(z,\lambda)$ or $\widetilde L^\pm(z,\lambda)$: $t^{(*)}(z,\lambda)$ is twist invariant.
Moreover the exchange relations of $t^{(*)}(z,\lambda)$ with the Lax matrices $L^\pm$ or $\widetilde L^\pm$ are identical since $t^{(*)}(z,\lambda)$ trivially commutes with $g(\lambda)$ and $g(\lambda+\sigma_z)$.

\begin{rmk}\label{rmk:Lchap}
Formulae \eqref{eq:deftmn}--\eqref{eq:deftmnst} can be recast in terms of the $U_{q,p}(\widehat{gl}_2)_c$ generators
$\widetilde{L}^\pm(z) = L^\pm(z,\lambda) \, e^{\sigma_z\partial}$ \cite{Konno}. 
In this case, $t^{(*)}(z_1,\lambda)$ commute with $\widetilde{L}^\pm(z_2)$
thanks to the exchange relations \eqref{eq:echtL}--\eqref{eq:echtstL}.
In other words, at the critical levels $c=\pm 2$, the $t^{(*)}(z,\lambda)$ generators are central in $U_{q,p}(\widehat{gl}_2)_c$. 
It is therefore natural to ask whether they are connected with the quantum determinant introduced in \cite{Konno2}. 
Direct inspection indicates that they are independent of the quantum determinant, see also remark \ref{rmk:Lchapsuite}. 
This strengthens their characterization as extended centers at the respective critical values similarly to the quantum affine \cite{FR1995} and vertex elliptic cases \cite{AFR}.
\end{rmk}

\subsection{Proof of Theorem \ref{thmone}}
The proof is decomposed into two lemmas, followed by a final step.
\begin{lemma}\label{lem:p0}
In $\elpb{2}$ and for arbitrary values of $c$, the operators $t^{(*)}(z,\lambda)$ obey the following relations:
\begin{align}
t^{(*)}(z_1,\lambda) \, L^\pm_2(z_2,\lambda) \, e^{\sigma_z^{(2)}\partial} = \frac{1}{\mathscr{N}(z_1,z_2)} \; & L^\pm_2(z_2,\lambda) \, e^{\sigma_z^{(2)}\partial} \, \tr_1 \big( N_1(\lambda-\sigma_z^{(2)}) \, R_{21}(\alpha z_2/z_1,\lambda)^{-sl_1-sl_2} \, \nonumber \\
& e^{-\sigma_z^{(1)}\partial} \, Q_1(z_1,\lambda) \, e^{\sigma_z^{(1)}\partial}\, R_{12}(\beta z_1/z_2,\lambda)^{-sl_1-sl_2} \big) \,.
\end{align}
where $Q(z_1,\lambda) = L^\pm(q^{\mp c} z_1,\lambda)^{-1} \, L^\mp(z_1,\lambda)$ and
the values of $\mathscr{N}(z_1,z_2)$, $\alpha$ and $\beta$ are given by: \\
\begin{tabular}{llll}
\null\qquad &-- for $t(z_1,\lambda)$ and $L^+_2(z_2,\lambda)$: 
&$\mathscr{N}(z_1,z_2) = \mathfrak{n}(q^{-c} z_2/z_1)$,\quad &$\alpha = \beta = q^{c}$, \\
&-- for $t(z_1,\lambda) $ and $ L^-_2(z_2,\lambda)$:
&$\mathscr{N}(z_1,z_2) = \mathfrak{n}(q^{2c} z_2/z_1)$, &$\alpha = q^{2c}$, $\beta = 1$, \\
&-- for $t^*(z_1,\lambda) $ and $ L^+_2(z_2,\lambda)$: 
&$\mathscr{N}(z_1,z_2) = \mathfrak{n}(z_2/z_1)$, &$\alpha = q^{-2c}$, $\beta = 1$, \\
&-- for $t^*(z_1,\lambda)$ and $L^-_2(z_2,\lambda)$: 
&$\mathscr{N}(z_1,z_2) = \mathfrak{n}(q^{-c} z_2/z_1)$, &$\alpha = \beta = q^{-c}$.
\end{tabular}
\end{lemma}
\proof
Let us start with the operator
\begin{equation}
t(z_1,\lambda) = \tr \big( N(\lambda) \, e^{-\sigma_z\partial} \, L^+(\gamma z_1,\lambda)^{-1} \, L^-(z_1,\lambda) \, e^{\sigma_z\partial} \big) \,,
\label{eq:t}
\end{equation}
where the matrix $N(\lambda) \in \End V \otimes \CC(\bar{\mathfrak{h}})$ is assumed to be diagonal and the value of $\gamma$ has to be determined. The explicit expression of $N(\lambda)$ leading to the desired result will be established in the course of the proof. \\
We first consider the exchange of $t(z_1,\lambda)$ with $L^+(z_2,\lambda)$. One has
\begin{align}
t(z_1,\lambda) \, L^+_2(z_2,\lambda) \, e^{\sigma_z^{(2)}\partial} &= \tr_1 \big( N_1(\lambda) \, e^{-\sigma_z^{(1)}\partial} \, L^+_1(\gamma z_1,\lambda)^{-1} \, L^-_1(z_1,\lambda) \, e^{\sigma_z^{(1)}\partial} \, L^+_2(z_2,\lambda) \, e^{\sigma_z^{(2)}\partial} \big) \nonumber \\
&= \tr_1 \big( N_1(\lambda) \, e^{-\sigma_z^{(1)}\partial} \, L^+_1(\gamma z_1,\lambda)^{-1} \, L^-_1(z_1,\lambda) \, L^+_2(z_2,\lambda+\sigma_z^{(1)}) \, e^{(\sigma_z^{(1)}+\sigma_z^{(2)})\partial} \big) \,.
\label{eq:tLm}
\end{align}
By the $RLL$ relations \eqref{eq:RLL} one gets
\begin{align}
t(z_1,\lambda) \, L^+_2(z_2,\lambda) \, e^{\sigma_z^{(2)}\partial} = &\tr_1 \big( N_1(\lambda) \, e^{-\sigma_z^{(1)}\partial} \, L^+_1(\gamma z_1,\lambda)^{-1} \, R^*_{21}(q^c z_2/z_1,\lambda+\sigma_z) \, L^+_2(z_2,\lambda) \nonumber \\
&\times L^-_1(z_1,\lambda+\sigma_z^{(2)}) \, R_{21}(q^{-c} z_2/z_1,\lambda)^{-1} \, e^{(\sigma_z^{(1)}+\sigma_z^{(2)})\partial} \big) \,.
\end{align}
A sufficient condition that allows the exchange of $L^+_1(\gamma z_1,\lambda)^{-1}$ and $L^+_2(z_2,\lambda)$ is to choose $\gamma=q^{-c}$. In this case one obtains
\begin{align}
t(z_1,\lambda) \, L^+_2(z_2,\lambda) \, e^{\sigma_z^{(2)}\partial} = &\tr_1 \big( N_1(\lambda) \, e^{-\sigma_z^{(1)}\partial} \, L^+_2(z_2,\lambda+\sigma_z^{(1)}) \, R_{21}(q^c z_2/z_1,\lambda) \, L^+_1(q^{-c} z_1,\lambda+\sigma_z^{(2)})^{-1} \nonumber \\
&\times L^-_1(z_1,\lambda+\sigma_z^{(2)}) \, R_{21}(q^{-c} z_2/z_1,\lambda)^{-1} \, e^{(\sigma_z^{(1)}+\sigma_z^{(2)})\partial} \big) \,.
\end{align}
Moving the rightmost exponential inside the trace to the left, it follows
\begin{align}
t(z_1,\lambda) \, L^+_2(z_2,\lambda) \, e^{\sigma_z^{(2)}\partial} = & L^+_2(z_2,\lambda) \, \tr_1 \big( N_1(\lambda) \, e^{-\sigma_z^{(1)}\partial} \, R_{21}(q^c z_2/z_1,\lambda) \, e^{\sigma_z^{(2)}\partial} \, L^+_1(q^{-c} z_1,\lambda)^{-1} \nonumber \\
&\times L^-_1(z_1,\lambda) \, e^{\sigma_z^{(1)}\partial}\, R_{21}^{-1}(q^{-c} z_2/z_1,\lambda)^{-sl_1-sl_2} \big) \,.
\end{align}
The properties of the $R$ matrix ensures that
\begin{equation}
e^{-\sigma_z^{(1)}\partial} \, R_{21}(x,\lambda) \, e^{\sigma_z^{(2)}\partial} = 
e^{\sigma_z^{(2)}\partial} \, R_{21}(x,\lambda)^{-sl_1-sl_2} \, e^{-\sigma_z^{(1)}\partial} \,.
\end{equation}
Hence one gets, after pushing the exponential $e^{\sigma_z^{(2)}\partial}$ outside the trace,
\begin{align}
t(z_1,\lambda) \, L^+_2(z_2,\lambda) \, e^{\sigma_z^{(2)}\partial} = & L^+_2(z_2,\lambda) \, e^{\sigma_z^{(2)}\partial} \, \tr_1 \big( N_1(\lambda-\sigma_z^{(2)}) \, R_{21}(q^c z_2/z_1,\lambda)^{-sl_1-sl_2} \nonumber \\
&\times \, e^{-\sigma_z^{(1)}\partial} \, Q_1(z_1,\lambda) \, e^{\sigma_z^{(1)}\partial}\, R_{21}^{-1}(q^{-c} z_2/z_1,\lambda)^{-sl_1-sl_2} \big) \,.
\end{align}
Finally, using the unitarity property of the $R$-matrix, one obtains
\begin{align}
t(z_1,\lambda) \, L^+_2(z_2,\lambda) \, e^{\sigma_z^{(2)}\partial} = \frac{1}{\mathfrak{n}(q^{-c} z_2/z_1)} \; & L^+_2(z_2,\lambda) \, e^{\sigma_z^{(2)}\partial} \, \tr_1 \big( N_1(\lambda-\sigma_z^{(2)}) \, R_{21}(q^c z_2/z_1,\lambda)^{-sl_1-sl_2} \nonumber \\
&\times e^{-\sigma_z^{(1)}\partial} \, Q_1(z_1,\lambda) \, e^{\sigma_z^{(1)}\partial}\, R_{12}(q^c z_1/z_2,\lambda)^{-sl_1-sl_2} \big) \,.
\end{align}
The other cases follow the same lines.\qed

We also need the following lemma:
\begin{lemma}\label{lem:p1}
Let $A_{12}$ and $C_{12}$ be elements of $\End V \otimes \End V \otimes \CC(\bar{\mathfrak{h}})$ and $M \in \End V \otimes \elpb{2}$. Then one has
\begin{equation}
\tr_1 \left( A_{12} \, e^{-\sigma_z^{(1)}\partial} \, M_1 \, e^{\sigma_z^{(1)}\partial}  \, C_{12} \right) = \tr_1 \left( (C_{12}^{sl_1.t_2} \, A_{12}^{sc_1.t_2})^{-sc_1} e^{-\sigma_z^{(1)}\partial} \, M_1 \, e^{\sigma_z^{(1)}\partial} \right)^{t_2} \,.
\end{equation}
\end{lemma}
\proof Direct calculation in components, i.e. through a projection of the above expressions on elementary matrices $e_{ij}\otimes e_{kl}$, followed by performing the trace in space 1.
\qed
\subsubsection*{Final step for the proof of theorem \ref{thmone}}
Applying lemma \ref{lem:p1}
to $A_{12} = N_1(\lambda-\sigma_z^{(2)}) \, R_{21}(\alpha \dfrac{z_2}{z_1},\lambda)^{-sl_1-sl_2}$, $C_{12} = R_{12}(\beta \dfrac{z_1}{z_2},\lambda)^{-sl_1-sl_2}$ and $M = Q(z_1,\lambda) = {L^{\pm}(q^{\mp c}z_1,\lambda)}^{-1} \, L^{\mp}(z_1,\lambda)$, one gets
\begin{align}
& \tr_1 \Big( N_1(\lambda-\sigma_z^{(2)}) \, R_{21}(\alpha \frac{z_2}{z_1},\lambda)^{-sl_1-sl_2} \, 
e^{-\sigma_z^{(1)}\partial} \, Q_1(z_1,\lambda) \, e^{\sigma_z^{(1)}\partial}\, R_{12}(\beta \frac{z_1}{z_2},\lambda)^{-sl_1-sl_2} \Big) 
\nonumber \\
& = \tr_1\! \left[ \left( \big[ R_{12}(\beta \frac{z_1}{z_2},\lambda)^{-sl_2}\big]^{t_2} 
 \big[N_1(\lambda-\sigma_z^{(2)}) R_{21}(\alpha \frac{z_2}{z_1},\lambda)^{-sl_1-sl_2} \big]^{sc_1.t_2} \right)^{-sc_1} 
\! e^{-\sigma_z^{(1)}\partial} Q_1(z_1,\lambda) \, e^{\sigma_z^{(1)}\partial} \right]^{t_2} 
\nonumber \\
& = \tr_1 \Bigg( \left[ \left( \big[ R_{12}(\beta \frac{z_1}{z_2},\lambda)^{-sl_2}\big]^{t_2} \, \big[N_1(\lambda-\sigma_z^{(2)}) \, R_{21}(\alpha \frac{z_2}{z_1},\lambda)^{-sl_1-sl_2} \big]^{sc_1.t_2} \right)^{-sc_1} \, N_1(\lambda)^{-1} \right]^{t_2} \nonumber \\
& \hspace*{280pt} \times N_1(\lambda) \, e^{-\sigma_z^{(1)}\partial} \, Q_1(z_1,\lambda) \, e^{\sigma_z^{(1)}\partial} 
\Bigg) \,.
\label{eq:formultrace}
\end{align}
A sufficient condition under which the RHS of \eqref{eq:formultrace} gives back the  $t^{(*)}$ generating function is that the $t_2$-transposed bracket in the trace be proportional to the identity matrix, i.e.
\begin{equation}
\left( \big( R_{12}(\beta \frac{z_1}{z_2},\lambda)^{-sl_2}\big)^{t_2} \, \big(N_1(\lambda-\sigma_z^{(2)}) \, R_{21}(\alpha \frac{z_2}{z_1},\lambda)^{-sl_1-sl_2} \big)^{sc_1.t_2} \right)^{-sc_1} \, N_1(\lambda)^{-1} = a(z_1,z_2) \II \,,
\end{equation}
for some non-vanishing function $a(z_1,z_2)$ (determined below). This last relation can be rewritten as
\begin{equation}
\Big(N_1(\lambda-\sigma_z^{(2)}) \, R_{21}(\alpha \frac{z_2}{z_1},\lambda)^{-sl_1-sl_2} \Big)^{sc_1.t_1} \, \big( R_{12}(\beta \frac{z_1}{z_2},\lambda)^{-sl_2}\big)^{t_1} = a(z_1,z_2) N_1(\lambda)^{sc_1} \,.
\end{equation}
Now, computing the first term of the LHS in components and taking into account the zero-weight property of the $R$-matrix, one gets
\begin{equation}
\Big(N_1(\lambda-\sigma_z^{(2)}) \, R_{21}(\alpha \frac{z_2}{z_1},\lambda)^{-sl_1-sl_2} \Big)^{sc_1.t_1} 
= \big( R_{21}(\alpha \frac{z_2}{z_1},\lambda)^{t_1}\big)^{-sc_2} \, N_1(\lambda-\sigma_z^{(2)})^{sc_1} \,,
\end{equation}
from which it follows
\begin{equation}
\left( \big( R_{12}(\beta \frac{z_1}{z_2},\lambda)^{-sl_2}\big)^{t_1} \right)^{-1} = \frac{1}{a(z_1,z_2)} \, \big( N_1(\lambda)^{sc_1} \big)^{-1} \, \big( R_{21}(\alpha \frac{z_2}{z_1},\lambda)^{t_1}\big)^{-sc_2} \, N_1(\lambda-\sigma_z^{(2)})^{sc_1} \,.
\label{eq:magic}
\end{equation}
Using corollary \ref{cor22}, equation \eqref{eq:magic} is satisfied when one chooses $\alpha \beta = q^{-4}$ and $N(\lambda) = G(\lambda)^{-sc}$, the function $a(z_1,z_2)$ being equal to $\mathfrak{n}(\alpha z_2/z_1)$ (or equivalently to $\mathfrak{n}(\beta z_1/z_2)$).

It remains to compare the relation $\alpha \beta = q^{-4}$  with the values obtained from lemma \ref{lem:p0}.
In the case of $t(z_1)$ and $L^\pm(z_2,\lambda)$, $\alpha\beta=q^{2c}$, therefore the condition is satisfied only when the central charge has the critical value $c=-2$. 
At the same time, in the case of $t^*(z_1)$ and $L^\pm(z_2,\lambda)$, $\alpha\beta=q^{-2c}$, 
hence the condition is now satisfied when $c=2$. For these specific values of the central charge, one finally obtains
\begin{equation}
t^{(*)}(z_1,\lambda) \, L^\pm_2(z_2,\lambda) \, e^{\sigma_z^{(2)}\partial} = \frac{a(z_1,z_2)}{\mathscr{N}(z_1,z_2)} \; L^\pm_2(z_2,\lambda) \, e^{\sigma_z^{(2)}\partial} \, t^{(*)}(z_1,\lambda) \,.
\end{equation}
Given the $q^4$-periodicity of the function $\mathfrak{n}(z)$, one finds that $a(z_1,z_2)=\mathscr{N}(z_1,z_2)$ for all cases.

Equations \eqref{eq:echtL}--\eqref{eq:echtstL} immediately follow.

\subsection{Vertex-IRF correspondence}\label{sect:VIRF}

The Vertex-IRF correspondence \cite{Baxter73,JMO} allows one to relate the Belavin--Baxter solution of the Yang--Baxter equation (which reduces to the 8-vertex $R$-matrix $R^{8V}(z_{12};p)$ in the $\widehat{gl}_2$ case) to the IRF matrix solution $R^{IRF}(z_1/z_2;p,w)$ of the Dynamical Yang--Baxter equation.
The Vertex-IRF transformation is implemented by a matrix $S(z;p,w)$ such that \cite{JMO,BRT,Yagi}
\begin{equation}
S_1(z_1;p,w) \, S_2(z_2;p,wq^{h_1}) \, R^{IRF}(z_1/z_2;p,w) = R^{8V}((z_1/z_2)^{1/2};p) \, S_2(z_2;p,w) \, S_1(z_1;p,wq^{h_2}) \,.
\label{eq:VIRFmat}
\end{equation}
Note that the spectral parameter dependence in \eqref{eq:VIRFmat} is chosen in order to connect directly to the $\elpa{2}$ $R$-matrix in \cite{AFR}.

The Vertex-IRF correspondence between the $R$-matrices allows one to define a morphism from $\elpb{2}$ to $\elpa{2}$:
\begin{equation}
\phi \;:\; L^{IRF}(z) \; \to \; S(z;p,w)^{-1} \, L^{8V}(z^{1/2}) \, S(z;p^*,wq^h) \,.
\label{eq:VIRF}
\end{equation}
Indeed, it is easily seen that if $L^{V8}(z)$ satisfies the vertex elliptic quantum algebra exchange relations \eqref{eq:RLLAqp}, then the combination $S(z;p,w)^{-1} \, L^{8V}(z^{1/2}) \, S(z;p^*,wq^h)$
obeys the dynamical exchange relations for $L^+(z,\lambda)$, see Eq. \eqref{eq:RLL}.
 
Consider now the trace formulae in \cite{AFR} given by
\begin{eqnarray}
t_{mn}(z) &=& \tr \Big( (g^{\frac{1}{2}} h g^{\frac{1}{2}})^{-m}\, L^{8V}\big((-p^{*\frac{1}{2}})^{n}z\big)\,
 (g^{\frac{1}{2}} h g^{\frac{1}{2}})^{-n}\, L^{8V}(z)^{-1} \Big) ,
\\
t^*_{-n,-m}(z) &=& \tr \Big( (g^{\frac{1}{2}} h g^{\frac{1}{2}})^{n}\, L^{8V}\big((-p^{\frac{1}{2}})^{-m}z\big)^{-1}\,
 (g^{\frac{1}{2}} h g^{\frac{1}{2}})^{m}\, L^{8V}(z) \Big) .
\end{eqnarray}
They satisfy the exchange relations 
\begin{eqnarray}
t_{mn}(z_1) \, L^{8V}(z_2) &=& \mathcal{F}_{mn}(z_1/z_2) \; L^{8V}(z_2) \, t_{mn}(z_1), 
\label{eq:exch1}
\\
L^{8V}(z_2) \, t^*_{-n,-m}(z_1) &=& \mathcal{F}_{mn}(z_1/z_2) \; t^*_{-n,-m}(z_1) \, L^{8V}(z_2), 
\label{eq:exch2}
\end{eqnarray}
with the original Lax matrix $L^{8V}(z)$ on the surface $\mathscr{S}_{mn}$ defined by $(-p^{\frac{1}{2}})^{m} (-p^{*\frac{1}{2}})^{n} = q^{-2}$.
Applying the pullback $\phi^*$ to \eqref{eq:exch1}--\eqref{eq:exch2} potentially yields a ``dynamically deformed'' exchange algebra:
\begin{equation}
s_{mn}(z_1;w) \, L^{IRF}(z_2) = \mathcal{F}_{mn}((z_1/z_2)^{1/2}) \; L^{IRF}(z_2) \, \widetilde{s}_{mn}(z_1;w,h)
\label{eq:lookslike}
\end{equation}
with
\begin{eqnarray}
s_{mn}(z;w) &=& S(z;p,w)^{-1} \, \phi^*\big(t_{mn}(z^{1/2})\big) \, S(z;p,w) \,, \\
\widetilde{s}_{mn}(z;w,h) &=& S(z;p^*,wq^h)^{-1} \, \phi^*\big(t_{mn}(z^{1/2})\big) \, S(z;p^*,wq^h) \,.
\end{eqnarray}
Although \eqref{eq:lookslike} looks close to \eqref{eq:echtL}-\eqref{eq:echtstL}, it is not enough to prove that we have a genuine dynamical exchange relation since a priori $\phi^*\big(t_{mn}(z^{1/2})\big)$ may depend on $w$, hence $\widetilde{s}_{mn}(z;w,h)$ would not be equal to $s_{mn}(z;wq^h)$. Solving this issue is beyond the scope of this paper. 

\begin{rmk}
\label{rmk:Lchapsuite}
We point out that applying the Vertex-IRF transformation to the object $t^*_{1,-1}(z)$ in $\elpa{2}$, using in addition the usual redefinition of $(L^{-})^{IRF}$ in terms of $(L^{+})^{IRF}$ \cite{JKOS,Konno}, the ratio of spectral parameters in the trace \eqref{eq:deftmn} is reproduced correctly.
Thus the potential mapping induced by $\phi^*$ suggests that our construction relates in the non dynamical case to the surface $\mathscr{S}_{1,-1}$. This explains why, as discussed in remark \ref{rmk:Lchap}, it does not reproduce a Liouville type formula, which corresponds in the vertex case to the surface $\mathscr{S}_{0,2}$. 
\end{rmk}

\subsection{Comments on the case $c=2$}\label{sect:comment}

A consistency issue arises from topological considerations, regarding the status of the generating functional $t^*(z)$ characterized by previous algebraic arguments as an extended (dyna\-mical) center for $c=2$. 
Indeed, if we consider as consistent the successive limits (see e.g. \cite{cladistics}) $\elpb{2} \to \mathcal{U}_{q}(\widehat{gl}_2)_c \to \widetilde{U}(\widehat{gl}_2)_c$ (assuming in particular that the specific nature of the first limit does not impair our conclusion), we are drawn to state that a suitable completion of the universal enveloping algebra $\widetilde{U}(\widehat{gl}_2)_c$ exhibits an extended center $\mathcal{Z}(\widehat{gl}_2)$ at $c=2$, contradicting the well-known theorem in \cite{FF1991} that extended center exists only at $c_{\text{crit}} = -h^\vee$ ($h^\vee$ is the dual Coxeter number of the Lie algebra). 
The issue here is of a topological, not purely algebraic, nature. The precise definition of the enveloping algebra entails indeed the construction of a local completion $\widetilde{U}(\widehat{gl}_2)_{c,loc}$ \cite{FF1991,BSch,FR1995} obtained inside an inverse limit $\underset{\longleftarrow}{\lim} \; \widetilde{U}(\widehat{gl}_2)_c / \widetilde{U}(\widehat{gl}_2)_c \big(gl_2 \otimes t^n \CC[t]\big)$ with $n>0$.
The local completion follows from the choice of a vacuum representation for the Kac--Moody algebra, where the vacuum vector $|0\rangle$ is annihilated by the positive modes of the generators. 
It is then linearly generated by Fourier coefficients of normal-ordered polynomials in the local fields, where of course normal-ordering is consistently defined w.r.t. the original vacuum. 
The elements of this completion are thus polynomials in the fields admitting some convergent normal ordering procedure
(by Wick's theorem or generalization thereof from a Poincar\'e--Birkhoff--Witt-type procedure).
It now follows that the generating functionals in the affine limit are such that
\begin{equation}
t(z) \in \widetilde{U}(\widehat{gl}_2)_{c,loc} \,, \qquad 
t^*(z) \notin \widetilde{U}(\widehat{gl}_2)_{c,loc} \,,
\label{eq:uloc}
\end{equation}
The proof of this statement follows from a direct inspection of formulae \eqref{eq:deftmn}--\eqref{eq:deftmnst} where the generator $L^+$ stands explicitly to the left or to the right of $L^-$. The limit to $\mathcal{U}_{q}(\widehat{gl}_2)_c$ decouples the positive modes in $L^+$ and the negative modes in $L^-$. From equation (4.42) in \cite{FRCMP146} one then sees that terms in any component of $t^*(z)$ are anti-normal ordered with the generic form $t^*_k \simeq \sum_{n>k} K_{ab} \, J^{a}_{n} \, J^{b}_{k-n}$ where the Kac--Moody algebra generators are denoted by $J^a_n$. Hence they cannot be brought to a normal-ordered form by any finite addition of normal-ordered terms: indeed the Wick ordering counterterm is divergent. By contrast all components of $t(z)$ are immediately normal-ordered in the sense of the vacuum defining $\widetilde{U}(\widehat{gl}_2)_{c,loc}$, hence \eqref{eq:uloc}.

The functional $t^*(z)$ exists in a mirror completion of the enveloping algebra $\widetilde{U}^*(\widehat{gl}_2)_{k,loc}$ based on a mirror vacuum representation  built from a vacuum $|0^*\rangle$ annihilated now by negative modes and using the opposite normal ordering prescription. 
The manifest algebra morphism $\psi : \widetilde{U}(\widehat{gl}_2)_{c,loc} \to \widetilde{U}^*(\widehat{gl}_2)_{-c,loc}$ where $\psi(a \otimes t^n) = a \otimes t^{-n}$ explains the occurrence of $c=+h^\vee$ as critical charge.
The drawback is that we expect $\widetilde{U}^*(\widehat{gl}_2)_{-c,loc}$ to have only lowest-weight representations.

The situation is expected to be similar for the extended centers in $\mathcal{U}_{q}(\widehat{gl}_2)_c$ since the corresponding modules have consistent deformations for $q$ generic. 
It is not so clear for the case of $\elpb{2}$ where a precise statement is still lacking regarding which completion of the enveloping algebra is to be used, as far as we are aware. 
Based on general categorical morphism arguments, we nevertheless conjecture that this issue may also arise.

\section{Conclusion\label{sect:conclu}}

Keeping in mind the topological issues discussed in section \ref{sect:comment}, we have established the existence of two quadratic trace-like combinations $t(z,\lambda)$, $t^*(z,\lambda)$ of generators of the dynamical elliptic quantum algebra $\elpb{2}$. 
We insist that the explicit form, although the trace contains shift operators $\exp(\pm\sigma_z\partial)$, lies indeed in $\elpb{2}$.
These two combinations both generate, respectively for values of the central charge $c=\pm 2$, an abelian subalgebra in a suitable completion of $\elpb{2}$. 
The generating functionals $t^{(*)}(z,\lambda)$ realize inside the algebra itself a ``dynamical center'' in the sense that they quasi-commute with the generators of $\elpb{2}$ encapsulated in quantum Lax matrices $L^\pm(z,\lambda)$ as \eqref{eq:echtL}--\eqref{eq:echtstL}.
At this stage, we interpret this ``dynamical commutation'' as an exact commutation in $U_{q,p}(\widehat{gl}_2)_c$. 

It is natural to ask for possible extensions to higher rank algebras $\elpb{N}$. 
The existence of a crossing-unitarity relation valid for all $gl(N)$ \cite{JMO} may be sufficient for such a purpose. However, in addition to this formula, a suitable splitting of the function $G(\lambda)$ is used. In the case $N=2$, such a splitting is a consequence of the crossing relation in \cite{Filali}. Unfortunately, the latter does not seem to admit easy extensions to $N > 2$. We hope to be able to address the problem of higher rank algebras in a forthcoming publication.
The case of $\elpbg$, where $\fg$ is any affine Lie algebra will be then a natural second step, initiated by the determination of a crossing relation for these algebras.

Commutation of $t^{(*)}(z,\lambda)$ with $t^{(*)}(w,\lambda)$ at $c = \pm 2$ suggests the possibility of deriving Poisson bracket structures by expanding the respective exchange relations around $c = \pm 2$. 
We expect that a similar structure as in \cite{ENSLAPP644} (although more intricate due to the dynamical shift) will arise to express the product $t^{(*)}(z,\lambda) \, t^{(*)}(w,\lambda)$ as a bilinear in the Lax matrix elements with a four-index kernel essentially obtained as a product of four $R$-matrices. 
A complete derivation of the Poisson bracket structures is left for further studies.

The difficulty in computing such structures is directly related to the final issue which we now raise.
We have only identified at this stage distinguished values of $c$ where the quadratic trace becomes at once dynamically central and abelian. 
We have however not yet identified critical ``surfaces'' in terms of $p,q,c$ (as was the case for elliptic algebras $\elpa{N}$), where the trace-like operator would generate a non-abelian subalgebra (possibly dynamical). 
The Vertex-IRF mechanism however suggests that such critical surfaces may exist in the dynamical case.
The quasi-periodicity condition expressing \emph{vertex-type} elliptic $R$-matrices at half-period of the spectral parameter by a rotation, was crucial in getting the critical surfaces. 
Its avatar in the dynamical case remains to be determined.

\paragraph{Acknowledgements:} We wish to warmly thank A. Molev for numerous discussions and fruitful comments regarding in particular extended centers, local completions and normal ordering of universal enveloping algebras. We also thank M. Jimbo for pointing out reference \cite{JMO} about crossing symmetry for dynamical elliptic algebras. We are indebted to the first referee for his profitable questions, which have led in particular to the subsection \ref{sect:VIRF}.

\end{document}